\documentclass[twocolumn,amsmath,amssymb]{revtex4}

\usepackage{graphicx}
\usepackage{dcolumn}
\usepackage{bm}

\newcommand{\etal}{\textit{et al.}}
\newcommand{\rmi}{{\rm i}}
\newcommand{\rme}{{\rm e}}

\begin{document}

\title{Rectified momentum transport for a kicked Bose-Einstein Condensate}

\author{Mark Sadgrove$^{1,2}$}
\email{mark@ils.uec.ac.jp}
\author{Munekazu Horikoshi$^{1}$}
\author{Tetsuo Sekimura$^{1}$}
\author{Ken'ichi Nakagawa$^{1,2}$}%
\affiliation{$^{1}$Institue for Laser Science, The University of Electro Communications, Chofushi, Chofugaoka 1-5-1, Japan}
\affiliation{$^{2}$Japan Science and Technology Agency}

\date{\today}

\begin{abstract}
We report the experimental observation of rectified momentum transport for a Bose-Einstein Condensate 
kicked at the Talbot time (quantum resonance) by an optical standing wave. 
Atoms are initially prepared in a superposition of the 0 and
$-2\hbar k_l$ momentum states using an optical $\pi/2$ pulse. By changing the relative phase
of the superposed states, a momentum current in either direction along the standing wave
may be produced. We offer an interpretation based on matter wave interference, showing that
the observed effect is uniquely quantum.
\end{abstract}

\pacs{Valid PACS appear here}
\maketitle

The current interest in rectified atomic diffusion, or atomic ratchets, may be traced
back to fundamental thermodynamical concerns~\cite{Feynman}
and also the desire to understand the so-called ``Brownian motors'' linked to 
directed diffusion on a molecular scale~\cite{Astumian,Riemann}. Abstractly, the ratchet effect may be
defined as the inducement of directed diffusion 
in a system subject to unbiased perturbations due to a broken spatio-temporal symmetry.

Given the scale on which such microscopic ratchets must work, 
it is not surprising that the concept of \textit{quantum ratchets}
has recently augmented this area of investigation.
The addition of quantum effects such as tunneling gives rise to new
ratchet phenomena such as current reversal~\cite{Riemann_PRL_1997}. Whilst early quantum ratchet investigations, 
both theoretical and experimental, have focussed on the role of dissipative
fluctuations in driving a ratchet current~\cite{dissapativeratchetrefs}, recent 
theory has considered the possibility of \textit{Hamiltonian ratchets},
where the diffusion arises from Hamiltonian chaos rather than stochastic fluctuations
\cite{HamiltonianRatchetRefs}.  This has lead to proposals~\cite{AORatchetTheory,BECRatchet}
and even an experimental realisation~\cite{Jones} for ratchet systems realised using
atom optics, in the context of the \textit{atom optics kicked rotor}~\cite{Raizen} where
periodic pulses from an optical standing wave kick atoms into different momentum states.

It is generally accepted that a ratchet effect cannot be produced without breaking the spatio-temporal
symmetry of the kicked rotor system. In Ref.~\cite{Jones},
a rocking sine wave potential was combined with broken time symmetry of the kicking pulses
to effectively realise such a system in an experiment. Other schemes
involve the use of quantum resonance (QR) to drive the ratchet effect.
At QR, atoms typically 
exhibit linear momentum growth \textit{symmetrical} about the initial mean momentum. 
However it has been suggested
that merely breaking the spatial symmetry of the kicked rotor at QR may be sufficient to produce
a ratchet current~\cite{QuantResRatchets}. 
\begin{figure}
\centering
\includegraphics[height=7.5cm]{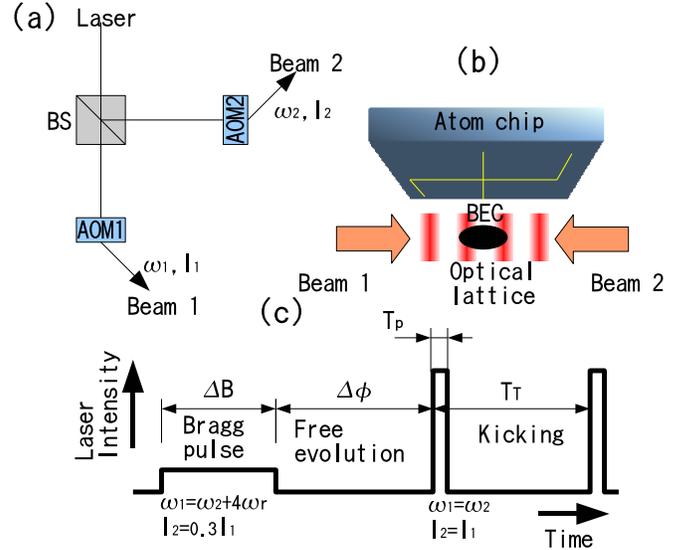}
\caption{\label{fig:expt} Diagrams of the experimental setup and sequence.
In (a), the laser configuration used to control Bragg diffraction and the 
kicking beam is shown. The beam is split by a 50/50 beam splitter (BS) 
and the output light passes through separate acousto
optic modulators (AOM) which control the beam intensities $I_{1,2}$ and frequencies $\omega_{1,2}$. 
The atom chip BEC setup is shown 
schematically in (b) along with the optical lattice created by the two intersecting
beams. The three different phases of Bragg diffraction, 
phase evolution and kicking are shown in (c) and are explained further in the text. }
\end{figure}
In this letter we present the first experimental evidence of such a resonant ratchet effect
in which the underlying mechanism is \textit{purely quantum}. Our system uses a Bose-Einstein 
condensate (BEC)
kicked by an optical standing wave~\cite{KickedBECrefs}, but there is no asymmetry in
either the kicking potential
or the 
period of 
the kicks, (which is set to the Talbot time $T_T$ corresponding to quantum resonance~\cite
{NistBEC}).
 Rather, the observed directed
diffusion is a property of the initial atomic wavefunction (which we prepare before kicking)
in the presence of a resonantly pulsed optical lattice. The experiment \emph{cannot} be  performed 
with thermal atoms, as it requires an initial atomic momentum spread much less than a photon recoil
in order to avoid dephasing effects. Our work presents
analytical, simulation and experimental results for a specific realisation of 
a ratchet at QR similar to that proposed in~\cite{QuantResRatchets}. We also offer 
a clear physical interpretation in terms of matter-wave
interference. 

As shown in Fig.~\ref{fig:expt}, our experiment is comprised of a BEC which is subjected
to pulses from an optical standing wave. The experimental configuration has been
explained elsewhere~\cite{HoriPRE,HoriAPB} and thus we provide only a summary here.
A BEC of $\sim 3\times10^3$ ${}^{87}$Rb atoms is realised and loaded onto an
atom chip~\cite{HoriAPB}. The atoms are trapped in the $5S_{1/2}$,$F=2$,$m_f=2$
state by the magnetic field generated by the chip and sit $700\mu$m below the
chip surface. Typically, the axial trapping frequency for the BEC is
$\omega_z \approx 2\pi\times 17$Hz and the axial and radial Thomas-Fermi
radii are $d_z=17\mu$m and $d_\rho=3\mu$m respectively.
The BEC is prepared in an
initial superposition state using a Bragg $\pi/2$ pulse 
and then kicked using light from a diode laser. Fig.~\ref{fig:expt}(a) shows the 
configuration used to control the intensity and frequency of the two beams used to create
the Bragg and kicking pulses. A free running 100mW diode laser, red detuned 4GHz (i.e. $\lambda=780.233$nm)
from the 
${}^{87}$Rb $5^2S_{1/2} \rightarrow 5^2P_{3/2}$ transition, enters a  
50/50 beam splitter and the output beams are passed through separate acousto-optic modulators (AOMs)
to control their frequency and amplitude after which, they intersect with the BEC (Fig.~\ref{fig:expt}(b)). 
\begin{figure}
\centering
\includegraphics[height=7cm]{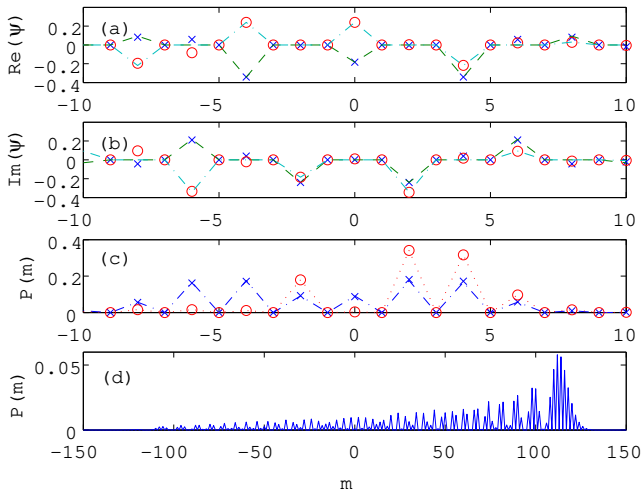}
\caption{\label{fig:wavefunc} Wavefunctions and momentum probability distributions
for kicked atoms with $K=0.6$. The momentum $m$ is in units of $2\hbar k_l$. In (a) and (b) respectively the real and imaginary parts  of
atomic wavefunctions after 5 kicks for $\phi=\pi$ are shown. The wavefunction evolving from
an initial $|0\hbar k_l\rangle$ state is shown with x's (simulations) and a dot-dash line
(theory of Eq.~\ref{eq:psiout}), whilst that which evolved from an initial 
$|-2\hbar k_l\rangle$ state is shown with $\circ$'s (simulations) and a dotted line (theory).
The lines are merely to guide the eye, and the theoretical wavefunction is only non-zero
at multiples of $m=2\hbar k_l$. In (c) asymmetry is seen to 
arise in the final momentum probability distribution  corresponding to $\phi=\pi$ 
(dashed line - theory, squares - simulations) whilst for $\phi=\pi/2$ (solid line - theory, 
triangles - simulations) 
there is symmetry about $m=-\hbar k_l$. In (d) the same system is shown 
after 100 kicks emphasising the extreme asymmetry of the momentum distribution.}
\end{figure}
The experimental sequence of laser pulses is shown in the diagram in Fig.~\ref{fig:expt}(c).

For the
Bragg pulse, the intensity of one beam is dropped to $3\%$ of its maximum power using the amplitude
modulation (AM) mode of one function generator whilst the 
frequency of the counterpropagating beam is increased by $4\omega_r\approx15kHz$ (where 
$\omega_r = 2.37\times10^4$Hz is the recoil frequency of ${}^{87}$Rb) relative to the other beam.
After the Bragg pulse, a period $\Delta_\phi$ of free evolution is used to adjust the quantum phase of the 
$|-2\hbar k_l\rangle$ state relative to $|0\hbar k_l\rangle$, and the beam intensity and
frequencies are made equal for kicking. The overall pulse envelope
and timing were controlled by another pulse generator.
The Bragg/kicking beams have an optical power of about 5mW. 
For a $\pi/2$ pulse, a duration $\Delta_B$ of $60\mu$s was used. For the kicking pulses,
a width of $T_p = 5\mu$s was used with a pulse period $T$ equal to the Talbot time
$T_T=\pi/2\omega_r\approx66.3\mu$s for ${}^{87}$Rb. 
\begin{figure*}
\centering
\includegraphics[height=9cm]{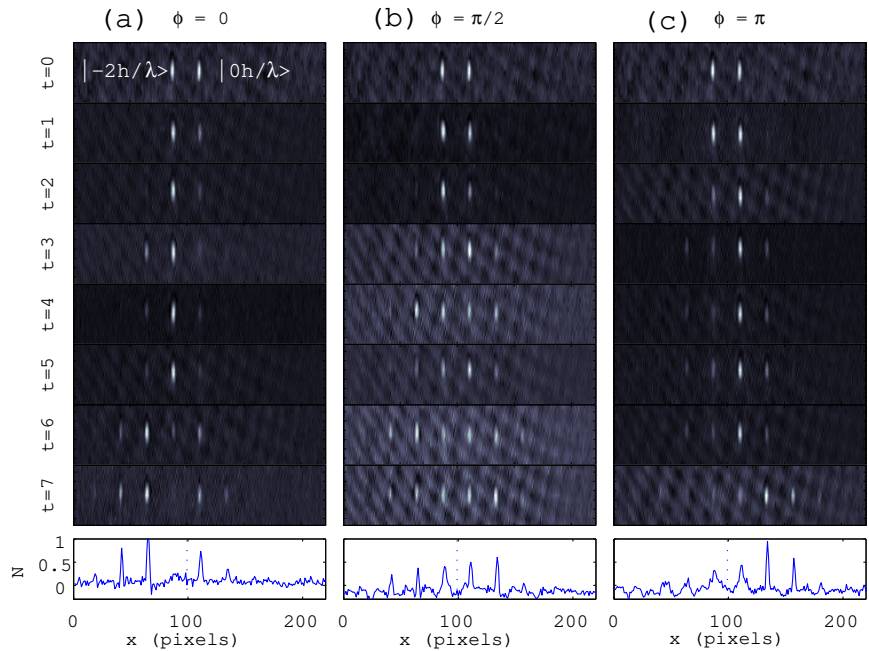}
\caption{\label{fig:slices}Sequences of absorption images for the ratchet
BEC experiment from $t=0$ to $t=7$ kicks for (a)$\phi=0$, (b)$\phi=\pi/2$ and 
(c)$\phi=\pi$. The $t=0$ case shows the initial distribution after the Bragg
$\pi/2$ pulse. In the top panel of (a), the 0 and $-2\hbar k$
momentum states are shown. 
At the bottom of each column, a sum over the rows of the image 
for $t=7$ is shown giving the distribution of atom number $N$ with position.
In these plots, the dashed line marks the position of the mean initial momentum
$\hbar k$. The images clearly show the presence of a ratchet current which reverses 
direction when the phase $\phi$ crosses $\pi=\phi/2$ (for which phase the current
is seen to vanish). }
\end{figure*}
Like other groups performing kicked BEC experiments~\cite{KickedBECrefs},
we have found that neither the energy due to atom-atom interactions nor the
harmonic potential affect our
results for the time scales used here, given the relatively much greater
energy due to kicking of the atoms. We simulate the system by
calculating the evolution of the initial wavefunction
subject to the single atom Hamiltonian \ref{eq:Ham}  (i.e. simulation of the Gross-Pitaevskii
equation is not neccessary). 

We now provide a theoretical treatment of our system. First we consider the preparation
of the initial state by a Bragg $\pi/2$ pulse. 
We will assume the the BEC starts in an initial 0 momentum eigenstate 
$|0\hbar k_l\rangle$. This is not a bad approximation, since the 
atoms in the BEC have a thermal spread which is much less than $2\hbar k_l$.
The $\pi/2$ pulse creates an equally weighted superposition state
$|\psi_B\rangle = \frac{1}{\sqrt{2}}\left(|0\hbar k_l\rangle -\rmi |2\hbar k_l\rangle\right)$.
After the Bragg pulse has been applied, a period $\Delta_\phi$ of free evolution is allowed.
During this time, the $|-2\hbar k_l\rangle$ state accumulates a phase 
$\phi=4\omega_r\Delta_\phi$, where 
$\phi=2\pi$ corresponds
to $\Delta_\phi=T_T$. The initial state just before kicking starts is then
\begin{equation}
\label{eq:psii}
|\psi_i\rangle = \frac{1}{\sqrt{2}}\left(|0\hbar k_l\rangle -\rmi\rme^{i\phi} |-2\hbar k_l\rangle\right).
\end{equation}
The dynamics due to sharp periodic momentum kicks applied to this state
are governed by the Hamiltonian~\cite{Graham}
\begin{equation}
\label{eq:Ham}
\hat{H} = \frac{\hat{p}^2}{2} + K\cos(2k_l\hat{x})\sum_{t}\delta(t'-t\tau),
\end{equation}
where $\hat{p}$ and $\hat{x}$ are the atomic momentum and position operators
respectively, $K=\hbar V_0T_p/\hbar$ is the kicking strength for an optical potential of
height $V_0$, $t'$ is time, $t$ is the kick 
counter and $\tau=4\pi T/T_T$ is the scaled kicking time.
The associated Floquet
operator for the case of QR ($\tau=4\pi$) is ~\cite{Shepelyansky}
$\hat{U}_{\rm q.r.}(t)=\exp(-{\rm i}Kt\cos(2k_l\hat{x}))$.
Applied to $|\psi_i\rangle$, the output wavefunction $\psi_o$ and momentum distribution $P(m)$
are~\cite{Cohen}
\begin{eqnarray}
\label{eq:psiout}
\psi_o(m)& = &\frac{\rme^{-i\frac{\pi}{2}m}}{\sqrt{2}}\left(J_m(Kt)-\rme^{i(\phi)}J_{m+1}(Kt)\right),\\
\label{eq:momdist}
P(m)  &=&\frac{1}{2}\left(J_m^2(Kt) + J_{m+1}^2(Kt) \nonumber \right.\\
& &\left. - 2\cos\phi J_m(Kt)J_{m+1}(Kt) \right).
\end{eqnarray}
Eq.s~\ref{eq:psiout} and \ref{eq:momdist} have a particularly interesting property: for general
phase $\phi$, the wavefunction and thus the momentum distribution \textit{grow asymmetrically 
with time}. This property is seen in Figs.~\ref{fig:wavefunc}(a) and (b) which show
 the wave function after 5 kicks  for $\phi=\pi$. The change in net momentum may be seen 
to be due to interference between the diffraction orders of the two initial wavefunctions 
which is mostly destructive below $m=-1$
but constructive above this initial mean momentum, leading to an asymmetric distribution of 
atoms (Fig.~\ref{fig:wavefunc}(c)). The dramatic nature of this induced asymmetry is demonstrated 
even more clearly in (Fig.~\ref{fig:wavefunc}(d)) which shows the theoretical probability distribution
after 100 kicks.  We note that the directed transport of atoms has been caused by the
interference of diffracted matter waves, that is, the observed ``ratchet'' effect is 
\textit{entirely quantum} 
(indeed, our experiment may be viewed as a type of atom interferometer~\cite{AtomIntRefs}).
Experimental confirmation is presented in Fig.~\ref{fig:slices}
which  shows absorption images of a kicked BEC after 
preparation into state $\psi_i$. The behaviour seen matches that predicted by 
Eq.~\ref{eq:momdist}. In particular, for $\phi=0$ the atomic momentum distribution
increases in asymmetry towards negative momentum, whereas for $\phi=\pi$, the 
asymmetry is in the opposite direction. For $\phi=\pi/2$  the 
distribution almost symmetrical (allowing for experimental fluctuations). 

We may also find the momentum current $i(t) = ({\rm d}/{\rm d}t)\langle p(t) \rangle$ by 
calculating the first 
moment of the momentum distribution 
$\langle p\rangle=\sum_m mP(m)=\frac{1}{2}\sum_m(mJ_m^2(Kt) + mJ_{m+1}^2(Kt) \nonumber
 - 2\cos(\phi) mJ_m(Kt)J_{m+1}(Kt))$. The first two terms 
give the momenta of the two superposed initial states e.g. 0 and -1 
(in $2\hbar k_l$ units)
respectively. The term of interest is $\sum_m mJ_m(Kt)J_{m+1}(Kt)$,
which may be summed by applying the standard Bessel recursion formula and the 
Neuman sum rule~\cite{BesselBook} to give $Kt/2$. Thus 
\begin{equation}
\label{eq:ratchetcurrent}
i(t) = \frac{\rm d}{{\rm d}t}\langle p(t) \rangle = -\cos\phi\frac{K}{2}.
\end{equation}

\begin{figure}
\centering
\includegraphics[height=7cm]{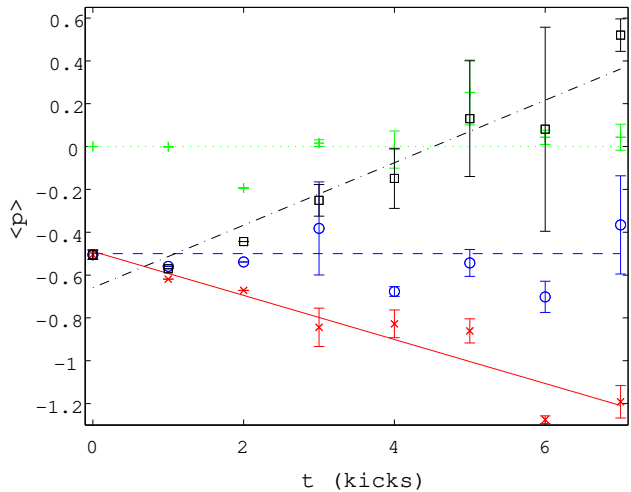}
\caption{\label{fig:current}The experimentally measured $\langle p\rangle$ (in units of $2\hbar k_l$) 
is shown along with 
theoretical curves for various initial conditions. Experimental data is shown by
x ($\phi=0$), $\circ$ ($\phi=\pi/2$), $\Box$ ($\phi=\pi$) and + (no initial $\pi/2$ pulse).
The solid and dash-dot lines are fits to the data for $\phi=0$ and $\phi=\pi$ respectively.
Dashed and dotted lines show $\langle p \rangle = -0.5$ and $\langle p \rangle = 0$ respectively
(note that these lines are not fits to the data).
}
\end{figure}

Eq.~\ref{eq:ratchetcurrent} offers a useful way to summarise the data.
The atomic momentum distribution was reconstructed from the absorption images shown in 
Fig.~\ref{fig:slices} and the mean momentum calculated. To check
repeatability we took another set of data for the same parameters as those in Fig.~\ref{fig:slices},
for $t>2$ (since very little diffusion  occurs in the first two kicks). Average currents for the two 
data sets are shown in Fig.~\ref{fig:current}, along with error
bars showing the difference between the measurements. The extraction of very small 
mean momenta ($\langle p\rangle\sim\hbar k_l$) from the ditributions in Fig.~\ref{fig:slices} is hampered by experimental imperfections 
such as CCD noise and scattered light, and laser frequency drift, which may lead to ocassional changes in
experimental parameters. This is the most likely cause of the large error bar seen in the 
case of $\phi=\pi$ when $t=6$. Increasing the accuracy of the measurements would require a larger atom
number and ideally a separate laser for Bragg diffraction and kicking. Nonetheless, Fig.~\ref{fig:current}
clearly demonstrates the momentum current effect and a current reversal for $\phi=0$ compared with
$\phi=\pi$. The data shows a general linear trend as predicted by Eq.~\ref{eq:ratchetcurrent},
with fitted lines shown in both cases. For $\phi=\pi/2$, although individual momentum distributions
are not perfectly symmetrical, the current is near 0 on average. The control
case for an initial $|0 \hbar k_l\rangle$ distribution is also shown and seen to exhibit 
near 0 average momentum current. Note that the dotted and dashed lines are not fits to the data, since there are no 
free parameters in either of these cases.
Theoretically, the momentum 
current should
persist indefinitely. In an experimental setting, however, imperfections such as the 
finite pulse width and any small difference between the pulse period and the Talbot time
will reduce the ratchet current. 
Due to a low signal to noise ratio at higher kick numbers in the current experiment, it was not possible to
probe these effects with our current setup.
We note that the effects seen here require a well defined
quantum phase between the initial states in superposition. Therefore, the experiment
must be performed using a BEC as a thermal cloud typically has a large spread of initial momenta
(and therefore quantum phase after free evolution), destroying the 
directed diffusion effect. It may be possible to exploit any sensitivity of the ratchet current 
to pulse-timing and phase variations to make accurate interferometry measurements.

In summary, we have demonstrated a novel quantum ``ratchet'' effect, in which 
directed momentum transport occurs in a system subject to a pulsed potential
with no net bias. The effect has no classical analogue, unlike previous such systems 
studied experimentally. 
The direction of the ratchet current varied with the initial quantum phase as predicted,
showing complete reversal for $\phi=0$ compared with $\phi=\pi$. This realisation of  
directed momentum transport suggests new possible mechanisms for directed motion on any scale
where quantum interference effects are non-negligible and resonant transport exists. 
M.S. would like to thank Scott Parkins and Andrew Daley for discussions regarding this work.

\end{document}